\documentclass[preprint]{aastex}
\shorttitle{Solar Helium Abundance}
\shortauthors{Laming \& Feldman}
\begin{document}
\title{The Solar Helium Abundance in the Outer Corona Determined from
Observations with SUMER/SOHO}
\author{J. M. Laming\altaffilmark{1} and U. Feldman\altaffilmark{2}}
\affil{$^1$ E. O. Hulburt Center for Space Research,\\Naval Research Laboratory Code 7674L, Washington DC 20375}
\email{jlaming@ssd5.nrl.navy.mil, tel: 202 767 4415, fax: 202 404 7997}
\affil{$^2$ E. O. Hulburt Center for Space Research,\\Naval Research Laboratory Code 7608, Washington DC 20375}
\email{ufeldman@ssd5.nrl.navy.mil, tel: 202 767 3286, fax: 202 404 7997}
\begin{abstract}
At altitudes of about 1.05 solar radii or more, the corona above quiet
solar regions becomes
essentially isothermal. This obviates many of the difficulties associated
with the inverse problem of determining emission measure distributions, and
allows for fairly straightforward relative element abundance measurements.
We present new values for the He abundance. The first is
based on a reanalysis of the
He/O ratio studied previously using data acquired by SUMER.  A more thorough evaluation of the
atomic physics for He II, including a detailed treatment of radiative recombination, increases
the predicted emission in the He II Balmer series compared to earlier analyses.
We use a recently revised value of the O abundance to
derive an He/H abundance ratio of 0.038, (mass fraction, $Y=0.13$), with an error
of $\sim 17\%$ coming mainly from the O abundance uncertainty. We demonstrate
that this result may be affected by gravitational settling of O relative to
He. We also derive an abundance for He by direct comparison to emission lines
of the H I Lyman series, with the result He/H $=0.052 \pm 0.005$ ($Y=0.17$),
a value similar to He abundances determined in the slow speed solar wind.

\end{abstract}

\section{Introduction}
The firm prediction of the primordial helium abundance has long been considered
one of the outstanding successes of big bang cosmology. It comes out in a
remarkably robust manner, from the neutron/proton ratio at weak interaction
freeze out, given by $n_n/n_p=exp\left(-Q/kT\right)\simeq 1/6$ where $Q$ is the
proton-neutron mass difference (in energy units), and $kT\sim 1$ MeV for
a radiation dominated universe. Almost all these neutrons end up bound in
$^4$He nuclei giving a big bang mass fraction $Y_{BB}=2n_n/\left(n_n +n_p\right)=
0.286$. Neutron $\beta$ decays before $^4$He formation reduce $n_n/n_p$ to
a value closer to 1/7, giving $Y_{BB}\simeq 0.25$. The main sensitivities
of $Y_{BB}$ are to the neutron half life (higher value gives weak interaction
freeze out at a higher temperature, and fewer neutron decays between freezeout
and $^4$He formation), and to the number of light particle species
(higher number
gives faster expansion rate for a given temperature, and hence weak freeze out
at higher temperature). This last sensitivity was used by \citet{steigmann77}
to place a limit on the number of lepton generations. Using
modern values for the primordial He abundance \citep[$Y_{BB}<0.243$][]{olive95}
and the baryon/photon ratio in the expressions of \citet{steigmann77}
gives $n_{\nu} < 4$, in accordance with experiments on the width of the
$Z_0$ resonance at CERN and SLC.

Helium is of course continually generated in stars as the end
product of reactions of the pp chain. Values for the primordial
helium abundance are best determined from metal poor extragalactic
H II regions, where recombination spectra of H I, He I, and He II
are visible. Extrapolations of He/H against O/H to zero abundance
for O yield mass fractions $Y_{BB}=0.232\pm 0.003 (1\sigma )\pm
0.005 ({\rm syst.})\le 0.243$ \citep{olive95}. \citet{sasselov95}
argue that a more careful treatment of the atomic physics and
radiation transfer for the observed recombination spectra could
increase $Y_{BB}$ to as much as $<0.255$. The solar system is
relatively young at $4.6\times 10^9$ years, and so the helium
abundance of the solar proto-nebula, $Y_{\odot }$, is expected to
be several times 0.01 higher than the primordial value, a value
consistent with measurements in the gas planets of the solar
system  \citep[c.f][]{niemann96,vonzahn96}. It was then something
of a surprise when values of the present day helium mass fraction,
$Y$, determined in the solar convection zone from helioseismology
(specifically from the discontinuity in the sound speed at the He
ionization zones, where the mean particle mass changes) came out
similar to the primordial value. Uncertainties in this technique
stem from the equation of state and the opacities used for the
solar envelope. For instance \citet{basu98} obtains $Y=0.248\pm
0.001$ and $Y=0.252\pm 0.001$ from OPAL and MHD models using data
from the SOI/MDI instrument on SOHO. \citet{kosovichev97} has
obtained $Y=0.248\pm 0.006$ and $Y=0.232\pm 0.006$ with the same
models respectively. These and other helioseismology uncertainties
are purely statistical coming from the data inversions. Systematic
uncertainties in thermodynamic parameters, e.g. the equation of
state, are not accounted for. These results were taken as evidence
that gravitational settling of He has occured within the solar
envelope, though note that these measurements are made in the
solar convection zone, where substantial gravitational settling
seems unlikely to occur. The proto-solar He abundance can be
determined from solar evolution calculations, including the He
settling in the solar radiation zone. The best match to current
solar properties (radius, mass, luminosity, etc) is achieved for a
$Y_{\odot }=0.273$ \citep{bahcall92}, although the inclusion of He
settling has a rather small effect on $Y_{\odot }$.
\citet{turck93} calculate a variety of solar evolution models
using different opacity databases and treatments of the effects of
electron screening on nuclear reaction rates and find $0.2661 \leq
Y_{\odot } \leq 0.2762$.

\section{The He Abundance in the Solar Upper Atmosphere (SUA)}
The first direct determination of He abundance from SUA plasmas
was done by \citet{gabriel95}. They used spectra recorded by the
CHASE instrument, which was launched into space aboard SPACELAB
II, to measure the intensity ratio between the H I Ly $\alpha$ I
1216 \AA\ and the  He II Ly $\alpha$ 304 \AA\  coronal lines, both
on the disk and at the limb. The spectrum they used was emitted by
a coronal quiet region that extended in height from 45000 to
135000 km (1-3 arcmin) above the limb. By assessing the resonantly
scattered components of these lines at the limb the He abundance
value they derived was $Y=0.23\pm 0.03$. Resonantly scattered He II 304 \AA\
radiation has also been identified in off-limb observations with the
Extreme Ultraviolet Imaging Telescope (EIT) on SOHO \citep{delaboudiniere99}.

\citet{raymond97} used a spectrum emitted by an equatorial
streamer at a height of $1.5R_{\odot}$ to determine the coronal He
abundance. The spectrum they used was recorded by the Ultraviolet
Coronagraph Spectrometer (UVCS) on SOHO. In their analysis they
compared intensity of the He II Ba $\gamma$ line at 1085 \AA\ with
the intensity of an H Ly line. Unfortunately, the He II Ba
$\gamma$ line was too faint to be detected in their spectra.
Therefore, they could only establish the value of 0.048 ($Y=0.16$)
as an upper limit for the He abundance. The upper value of the
abundance that \citet{raymond97} derived was only 60\% as high as
the He abundance value that was derived by \citet{gabriel95}.

In a study on the plasma properties of a quite sun streamer between
$1.03R_{\odot}$ and $1.5R_{\odot}$
that was based on SUMER spectra, \citet{feldman98b}
showed that elemental settling can greatly influence the composition of
SUA plasmas. As a result of this study \citet{feldman98b}
postulated that the low value obtained by
\citet{raymond97} was a result of elemental settling effects that
influence plasma compositions at large heights above the solar
surface. The same could perhaps to a smaller extent affect the \citet{gabriel95}
measurements.
To minimize the effect caused at large heights by
elemental settling, \citet{feldman98a} used a spectrum that was emitted by
plasma located at a height of $1.03R_{\odot}$. In order to improve the
statistics
of the line intensity ratio \citet{feldman98a}
chose to compare He II Ba $\gamma$
1085 \AA\  intensity with the intensity of the much brighter O VI
2s $^2{\rm S}_{1/2}$-2p $^2{\rm P}_{3/2}$ line at 1031 \AA\ . By using the same set of atomic
data as \citet{raymond97} he found $Y=0.25\pm 0.04$. This
result was revised in a preliminary version of this work \citep{laming99b} to
$Y=0.26\pm 0.04$.

The He abundance has also been measured in situ from particle measurements made
in the solar wind. The fast solar wind that is associated with coronal holes
has a quite constant He abundance of
4.8 \% relative to H ($Y=0.16$). The slow solar wind, associated with quiet regions,
has a slightly lower He fraction ($\sim 4$\%, $Y=0.14$),
but is quite variable \citep[see][]{feldman97,steiger97}.
Some of this
variability comes from high speed streams and their associated magnetic sector
boundaries, which can increase or decrease the He abundance respectively,
together with shorter term variations associated with transient
phenomena such as flares.
\citet{reames94} review a large number of these events.

\section{SUMER Observations and Analysis}
The SUMER observations analysed in this paper are described in
detail elsewhere \citep{feldman99}. Briefly, an equatorial
streamer was observed on 1996 November 21/22 during a spacecraft
roll maneuver, so that the SUMER slit was oriented in the
east-west direction, instead of the usual north-south direction.
We focussed on the data taken with the $4\times 300$ arcsec$^2$
slit centered at 1160 arsec from solar center, or 1.177 solar
radii. The data were flatfielded and destretched using standard
SUMER procedures.

The choice of lines with which to compare the He II emission requires some care.
Element abundances in the quiet corona are in general different to those in the
photosphere, in that elements with a First Ionization Potential (FIP) less
than about 11.5 eV are enriched by a factor of typically 4, relative to hydrogen,
whereas the so-called high FIP elements (with FIP greater than 10 eV) have
no enrichment at all. In coronal hole regions the FIP
effect does not appear to operate, and in active regions it
may cause intermediate enrichments between 1 and 4. In very old active regions
and polar plumes, enrichments of
up to an order of magnitude or more have been reported, and so the best
elements for comparison to He II are the high-FIP elements, where no such
behaviour has been recorded. These elements include N, O, Ne, Ar.

The most important feature of the abundance determinations of \citet{raymond97}
and \citet{feldman98a} is that at sufficiently high altitudes, the solar
corona becomes essentially isothermal. This is important since the He II
emission lines are very temperature sensitive at transition region
temperatures where their emissivity is maximized, but are much less so at
coronal temperatures. However the small fraction
of He that retains one electron at the temperature
in the more quiescent coronal structures
emits sufficient radiation to be measured and compared to other ions formed
at coronal temperatures, e.g. O VI.

This isothermal property is demonstrated in Figure 3 of
\citet{feldman98b}, which shows the apparent element abundance
enhancement over that of oxygen as a function of temperature in
the equatorial streamer (the same data as analyzed herein for the
He abundance) for high and for low FIP elements. In each case the
curves overlap close to $\log T=6.15$. In non-isothermal
conditions the various intersections between the curves for
different ions would appear at widely varying temperatures. A
similar plot using lines from various ions of Si from this same
data set can be found in Figure 4 of \citet{feldman99}, giving a
temperature of $\log T=6.11\pm 0.04$. We take this value as the
plasma temperature in the analysis that follows. We omitted the
curve due to the Li-like ion Si XII from our considerations, since
it lies at the extreme short wavelength end of the SUMER bandpass,
where calibration difficulties are suspected to exist
\citep{laming97}.

\subsection{Instrument Issues}
The He II Ba $\gamma$ multiplet consists of the 2s-5p 1084.913 \AA\  and 2p-5d
1084.975 \AA\  lines. The He II lines appear in close wavelength
proximity
but separated from several N II lines (1083.990, 1084.562, 1084.580,
1085,529, 1085.546, 1085.701 \AA\ ) that belong to the
2s$^2$2p$^2$ $^3{\rm P}$-2s2p$^{3~3}{\rm D}$
multiplet. The front mirror of the SUMER telescope is not perfectly
smooth. As a result the intensity of any line recorded above the limb
contains a genuine coronal component and a component due to scattered
light that originates from emission formed on the solar disk.
\citep[For a description of the scattering properties of the SUMER
instrument see][]{feldman99}. The He II lines that were
recorded
at a height of $1.03R_{\odot}$ above the quiet limb are no exception. Their
intensity needs to be de-convolved to the coronal and scattered
components. Figures 1 and 2 show fits to the spectral region around the
He II 1085 \AA\  multiplet. The four peaks fitted to the data are from
left to right the N II 1083.98, 1084.58 \AA\  transitions, the He II
1084.94 \AA\  Balmer $\gamma$ multiplet, and finally a blend of the N II
1085.53, 1085.55, and 1085.68 \AA\  lines. Figure 1 shows a fit to a spectrum
taken from pixels between 88 and 118 arcseconds from the solar limb, while
Figure 2 shows data from pixels between 243 and 273 arcseconds altitude. In
both cases the N II lines are dominated by scattered light. However the
increase in the intensity of the He II multiplet relative to the N II lines in
Figure 1 is due to the true coronal emission. The ratio between He II and N II
far from the solar limb in Figure 2 was used to assess the contribution of
scattered light to the He II observed close to the limb, and to subtract it
from the total emission leaving only the true coronal emission. Fitted
intensities for He II and the ratios of these to the total of N II emission
for seven positions off limb are given in Table 1, for each of the two detector
positions that recorded these data on the KBr portion of the photocathode. For
reference, the results of similar fits to the He II 992 region, giving the
intensities of the He II 992.39 \AA\  (Balmer $\epsilon$) multiplet and its
ratio to the nearby N III 991.57 \AA\  line are given in Table 2.
The measured intensity ratio He II 992/1085 is $0.265 \pm 0.28$, about $1.5\sigma$ away
from the predicted ratio of 0.315 (see discussion of He II intensities below). In view
of the radiative and collisional contributions to the excitation of H I discussed later
in this paper and in \citet{raymond97}, we
do not expect any radiative excitation component to the coronal He II emission, and so
the He II 992 \AA\  intensities are not used in further analysis.

\subsection{The O Abundance and Emission Lines}
At the height of $1.03R_{\odot}$ \citet{feldman98a} assumed that
the relative composition between He
and O is still photospheric, i.e., no elemental settling effects are as
yet at work, and
used a value for the oxygen abundance value of log(O)=8.93
\citep[taken from][]{anders89}. The CNO abundances have been
recently revised \citep{grevesse98}, giving a new value of log(O)=8.83;
a decrease of 26\%. Thus in comparison to this the He abundance of
\citet{feldman98a}
should be revised downwards to 0.063 by number relative to hydrogen,
giving a mass fraction $Y=0.21$. The uncertainty in this new oxygen abundance
is given by \citet{grevesse98} as 0.06 dex, or $\pm 15$\%. As we shall
see below, this is likely to be the limiting accuracy in determining the He
abundance relative to O.

The pixel range over which the  O VI line
intensities were integrated are given in the
first column of Table 3.
The intensities measured for the O VI 1031.93 and 1037.60 \AA\  lines are
given in the next two columns of Table 3. We assume that these lines are
optically thin high in the corona. The intensity ratio $R$ between them
is given in the fourth column. Under conditions of pure collisional excitation
this ratio should be equal to 2, and its increase from this value gives an
indication of how much of the O VI intensity is radiatively excited. The
fraction of the O VI 1031.93 \AA\  line that is radiatively excited is given
in the fifth column, and the sixth column gives intensity ratios for
He II 1085/O VI 1031, for the positions off limb where a meaningful value can
be derived. These ratios have been corrected for the slightly different
spectrometer sensitivity at the two wavelengths.
The apparent increase in the abundance ratio He/O with height
above the limb is immediately suggestive of gravitational settling, which would
deplete O relative to He. Since it is apparent that the SUMER
destretching procedures do not necessarily map exposures taken on different
detector positions to exactly the same pixels, we have decided to
leave out data taken from
the extreme ends of the detector. For example note the relative
intensities between He II 1085 \AA\  and 992 \AA\  at the two lowest altitudes
reported in tables 1 and 2. Another rationale is that the isothermal condition
of the solar corona necessary for our analysis is more likely to break down
at lower altitudes. As a result our lowest measured position was chosen to be
at 88-118 arcseconds, a position which is higher than the one measured by
\citet{feldman98a}. At this position, the He/O ratio is still higher than that
measured by \citet{feldman98a}.

\subsection{Atomic Physics Issues}
\citet{feldman98a} used the same atomic data for the O VI and He II transitions
as did \citet{raymond97}. While the O VI rates are straightforward and
uncontroversial, those for He II require more attention. Very few calculations
available treat the neccesary electron energy region in the excitation process
to be relevant to coronal He II emission. \citet{raymond97}
used He II collision strengths taken from
close-coupling approximation calculations \citep{mclaughlin96}. These calculations
only included levels up to $n=3$, and so rates were scaled according to the
absorption oscillator strengths to get results for the $n=5$ excitation. A
correction for the assumed population by cascades of +20\% was also included.
A particular subtlety of neutral or singly ionized atom excitation is that
in general for electron impact energies above the ionization threshold, the
ionization channel must be explicitly included in the calculation to get the
excitation cross sections right. The reason for this is that the sum of all
transition probablities cannot be greater than unity, but if a physically
significant process is left out of the calculation, this unitarity bound cannot
be guaranteed. Thus the He II data used by \citet{raymond97} for collisional
excitation are likely to
be overestimates of the true excitation rates.

We have recently compiled a model He II ion using electron impact cross section
data calculated by \citet{bray93}. These authors included a full treatment
of electron impact ionization, for excitation from the ground state to levels
up to $n=4$, and electron energies up to 700 eV. We scaled their rates for
$n=4$ by the absorption oscillator strength to get impact excitation rates going
up to $np$ states, and by $1/n^3$ for $ns$ and $nd$ states. Proton excitation
rates among the $n=2$ levels are taken from \citet{zygelman87}, and these
results are scaled for proton rates among $n=3$ and higher levels.
Radiative decay rates for E1 transitions were calculated from the
expression in terms of hypergeometric functions given in \citet{bethe57}, and
those for the 2E1 transition are taken from \citet{drake81}. We also include
radiative recombination into all states of He II from the bare charge state,
calculated using a subroutine due to D. G. Hummer \citep[c.f.][]{hummer87}.
This last process produces
substantial population in excited levels, and so we investigated various sizes
of model ion. We base our results (given in Table 4) on calculations including
levels up to $n=20$ in He II, i.e. 400 levels in total, plus one for the bare
charge state, giving a $401\times 401$ matrix to invert. The ionization rate between He II
and the bare charge state is taken from \citet{arnaud85}. Going up to $n=30$ (901 levels)
produced further corrections of order 0.5\% or less to calculated line emissivities. The
density is taken to be $10^8$ cm$^{-3}$, as determined from Figure 5. in
\citet{feldman99} from the Si VIII lines. Variations in density of up to a
factor of two (i.e. much larger than the uncertainty in the plot in this last
reference) produce changes in the theoretical emissivities of less than 1\%.

The inclusion of radiative recombination produces significant extra emission
in lines of the He II Balmer series, leading to an overall lower value of the He
abundances than derived in our earlier papers \citep{laming99b,feldman98a}, where
this process was not included in the analysis. Table 4 should also be considered
as superceding Table 6 in \citep{laming99a}. We checked our results against Case A
recombination spectra for a temperature of 30,000 K tabulated by \citet{clegg99} (this is
the highest temperature they consider). To
simulate a photoionization-recombination equilibrium the ionization rate in our runs at
this temperature was increased by a large factor to put essentially all the helium into
the bare charge state. Our results in this case were consistently around 6\% lower than
those of \citet{clegg99}, for a variety of emission lines and densities,
which presumably stems from different choices for atomic data.
Our new He II emissivities and those for O VI
given as a function of temperature by \citet{laming99b}, give an
intensity ratio for He II 1085\AA\ /O VI 1032\AA\
of $\left(1.59\pm 0.02\right)\times 10^{-3}$ for a
temperature of $\log T=6.11\pm 0.04$, using the O and He photospheric abundances
from \citet{grevesse98} (8.83 and 10.93 respectively).
When compared with the results in Table 3, this indicates He underabundant relative
to O by factors of $0.45\pm 0.04$, $0.47\pm 0.05$ and $0.54\pm 0.07$. The small decrease with
height of the above ratio is
consistent with a depletion of O relative to He by gravitational settling as
mentioned previously. Expressed relative to H the He abundance from the
lowest altitude studied is $0.038\pm 0.007$ where the 8\% measurement error
and the 15\% uncertainty in the O abundance have been added in quadrature.

\subsection{The He Abundance Measured Relative to H}
Lines of the H I Lyman series are also visible in our SUMER spectra. Far
from the limb these are dominated by scattered light from the disk, but for
the lowest altitude position consider (88 - 118 arcseconds)
some true coronal H I emission is detectable. We subtract off the
scattered light component using the same ratio determined from the N II
and N III lines in the He II fits, whereby the scattered light at position
88-118 arcseconds is a factor of $1.825\pm 4$\% more intense than that at
243-273 arcseconds.
\citet{feldman99} have shown that the gross features of the scattered light
are essentially independent of wavelength. However at the few per cent level
at which we need to work, we did see a variation in the
decrease in the scattered light moving away from the limb, with shorter
wavelengths (e.g. C III 977.02 \AA\ ) decreasing less fast than longer
wavelengths (e.g. Si III 1113.22 \AA\ ). We therefore found that insignificant
improvement in the subtraction of scattered light was to be found by
considering more lines than just the two most optimum N II and N III features
mentioned above.

From the relative intensities of the Lyman $\beta$ 1025.72 \AA\  and the
Lyman $\gamma$ 972.54 \AA\  we are able to calculate the contribution of
radiative excitation to each line. A model H I atom
identical in all respects to that for the He II ion was compiled using electron
impact data from \citet{bubelev95}. Proton rates for this model come from a
code described in \citet{laming96}, originally applied to the 2s-2p proton
excitation of Li-like ions. Results from this code for He II are in good
(i.e. $<20$\%) agreement with the results of \citet{zygelman87}. The ionization rate
between H I and the bare charge state is taken from \citet{scholtz91}. Again, in checks
against \citet{clegg99} at 30,000 K a consistent underestimate of around 6\% was found,
suggesting that in forming the ratio of emission He II/H I, these deviations from
\citet{clegg99} will cancel allowing us to compute a theoretical line ratio with very high
accuracy. Emissivities for the Lyman series of H I
from this model are given in Table 5, again for an electron density of
$10^8$ cm$^{-3}$. The He II emissivities are as independent of density
in this range as the H I results. At $\log T=6.11\pm 0.04$ the ratio of
Lyman $\gamma$/Lyman $\beta$ is predicted to be $0.336\pm 0.002$.
Taking the ratio of
disk intensities of these two lines from the scattered light observed between
243 and 273 arcseconds and scaling by
the absorption oscillator strengths, the intensity ratio in pure radiative
excitation would be 0.088. The observed ratio, after correcting for scattered
light, subtracting off the contributions from the He II Balmer multiplets
that are blended with the H I Lyman lines (calculated from the observed
intensities of the He II 1085 and 992 \AA\  multiplets), and applying the
correction for the wavelength dependence of the spectrometer sensitivity gives
an observed Lyman $\gamma$/Lyman $\beta$ intensity ratio of $0.248\pm 4$\% ,
where the uncertainty is dominated by the subtraction of scattered light. Thus
the collisional $c_x$ and radiative $r_x$ components of Lyman $\beta$ are
related by $c_x=r_x\times\left(0.088-0.248\right)/\left(0.248-0.336\right)
=1.82r_x$. Hence the fraction of Lyman $\beta$ that is collisionally excited
is $c_x/\left(c_x+r_x\right)=1.82/2.82=0.65$, with a probable error of
$\pm 0.04$. This reduction of the H I line intensities is summarized in
Table 6.

The intensity ratio
He II 1085/H I 1025.72 is then $0.036\pm 9.8$\%. The
theoretical ratio
based on an He abundance of 0.085 relative to H ($Y=0.25$) is $0.059\pm 0.3$\%,
where the error comes from the uncertainty in the temperature. This yields
a measurement of the abundance ratio He/H of $0.052\pm 10$\% by number or
a mass fraction of $Y=0.17\pm 10$\%. The uncertainty is dominated by the
counting statistics and subtraction of scattered light for the He II 1085 \AA\
multiplet, and by the determination of the fraction of H I Lyman $\beta$ that
is collisionally excited. The improvement of this last factor in particular
requires subtraction of scattered light to very high precision, and is likely
to be the limiting factor.

\section{Discussion and Conclusions}
Our two measurements of the abundance ratio He/H are affected in opposite
directions by gravitational settling (assuming that this affects ions in order
of their mass, with O being the most affected and H the least). There are also unknown
systematic errors associated with the O VI ionization balance (H I and He II should be
much more secure in this respect). Therefore we discuss the He abundance determined
relative to H of $0.052\pm 0.005$. This is
a 1 $\sigma$ uncertainty, and includes all known sources of error. Thus, barring any
problem with the atomic physics input, we may
claim to have the most precise measurement of the helium abundance in the solar
corona. The result is strikingly close to He abundance measurements in the slow
speed solar wind, especially bearing in mind that this He fraction ($\sim 4$\%)
is known to be quite variable. The reason why this result is lower than previous
estimates \citep{feldman98a,laming99b} from the same observations is the inclusion of
radiative recombination as a population mechanism in He II and H I. It has a stronger
effect on lines of the Balmer series, and so the He II 1085 \AA\  emissivity is increased
more than that for the H I 1025 \AA\ , leading to a decrease in the inferred He/H abundance
ratio. By including a similar correction to the \citet{raymond97} measurement an upper
limit of He/H $<0.21$ is derived.

Our result also agrees well with the modeled valued of the coronal He abundance
for steady-state conditions given by \citet{hansteen94}. These authors also
discuss time dependent solutions, finding enhancements in the coronal He
abundance when the coronal temperature and corresponding flux of slow solar
wind protons fall below certain limits. This occurs because the proton-He
coupling in the solar wind is sufficiently weak that the protons, as they flow
out into the solar wind, essentially leave the He behind in the corona. The
coronal temperature in our study, $\log T=6.11\pm 0.04$ is sufficiently low
that according to \citet{hansteen94}, this ought to be the case here. However
our result for the He abundance is much more consistent with the higher temperature
coronal models where \citet{hansteen94} find steady state solutions. Further work
\citep{hansteen97} has investigated the role of chromospheric mixing processes on the
coronal and solar wind He abundances.

Various other authors have suggested that the
helium abundance could vary in different solar regions. In a model to explain
the observed preferential acceleration of $^3$He in impulsive solar flares,
\citet{fisk78} found that the necessary electrostatic ion cyclotron waves
were more efficiently excited in plasma by electron drifts
where the He abundance was increased
from its usual value by a factor 2-3 or more. More recent work on this
phenomenon appears to have obviated the need for an increased He abundance
\citep[c.f.][and references therein]{steinacker97,miller98}
in the flare plasma by replacing the
electron drift in Fisk's work with a flare accelerated electron beam. In
a review of available measurements and models \citet{drake98} concluded that
He abundance enhancements on the order of a factor of 2 over photospheric
values could not be excluded
among different regions of solar and stellar coronae. This work was primarily
concerned with observations of thermal bremsstrahlung continua in stellar
coronae that appear to be too intense for a hydrogen dominated corona, where
an increase in the coronal He abundance would go some way to alleviating this
difficulty. \citet{share98}, studying gamma ray spectra observed by the Solar
Maximum Mission and the Compton Gamma ray Observatory,
found evidence for an enhanced He abundance relative
to protons present in flare accelerated particles. Such abundance enhancements
by factors of as much as
5 were noted, whereas ambient photospheric material with which the flare
accelerated particles interact was found to be adequately described by the
usual He abundance relative to H.
Of course there is also the possibility the He is depleted in the
corona, as evidenced by solar wind observations described above. We defer
further comment on this issue of the variability of the He abundance to a later
work, but emphasise that the precision of our measurement technique is more
than adequate to allow thorough tests of these ideas in various regions of the
solar corona.

The one aspect of the quiescent solar corona crucial to our work, that it is
essentially isothermal, has recently been questioned by \citet{wolfson00}.
Arguing from YOHKOH SXT
observations made of the same streamer at the same time as our
SUMER observations, these authors claim that the coronal temperature slowly
increases with height above about 1.13 solar radii. The precise nature of
their claim does not affect our results (the corona is still isothermal
at lower altitudes, where our He abundance measurements are made). Analysis of another
equatorial streamer observed on 1996 July 22-27 by both the Ultraviolet Coronagraph
Spectrometer (UVCS) on SOHO and the YOHKOH SXT does find isothermal conditions
\citep{li98}. For the streamer under study in this paper, we
still favour conclusions drawn from SUMER observations to
those from the YOHKOH SXT, since the SXT data consist of
ratios of count rates in two filters, the AlMg and Al filters,
not actual spectra, and the strongest lines
transmitted by these filters (e.g. O VII, O VIII, Fe XVII)
originate from higher temperatures
than those present in the equatorial streamer. We determine the temperature
from a range of Si ions \citep[see Figure 5 in][]{feldman99} whose temperatures
span a more relevant range for this purpose. Of course our abundance
determinations use ions usually formed at much lower temperatures than those
present in the streamer, but these are hydrogenic ions for which atomic
processes can be calculated with very high accuracy.

\acknowledgements
This work was supported by the NRL/ONR Solar Magnetism  and the Earth's
Environment 6.1 Research Option and by NASA Contract W19473. We are grateful
to John Raymond for  advice and encouragement, and to Igor Bray for making
his cross section data available to us in electronic form. The SUMER project
is financially supported by DARA, CNES, NASA and the ESA PRODEX program
(Swiss contribution). SUMER is a part of SOHO, the Solar and Heliospheric
Observatory, of ESA and NASA.

\begin{figure}
\caption{Fit to He II and N II features near 1085 \AA\  recorded at an altitude
between 88 and 118 arcseconds from the solar limb.
From the left the lines are the N II 1083.98 and 1084.58, the He II
1084.94 (i.e. the Balmer $\gamma$ multiplet), and the N II 1085.53, 1085.55
 and 1085.68
lines. The last three N II lines are blended together into one feature. The
solid histogram shows the data, the dotted line the fit, the dashed line the
residuals, and the dash-dot lines shows the individual components in the fit.}
\end{figure}

\begin{figure}
\caption{Fit to the same lines as in figure 1, but recorded between altitudes
of 243 and 273 arcseconds from the solar limb. The diminution in intensity
of the He II line (third from left) relative to the other N II lines from
that in figure 1 can clearly be seen, as the coronal He II emission disappears
and the spectrum is purely scattered light from the solar disk.}
\end{figure}


\begin{table}
\caption{He II 1085 \AA\ Intensities. \label{tab1}}
\begin{tabular}{rrrrr}
\\
position\tablenotemark{a} & He II\tablenotemark{b}&
He II/N II\tablenotemark{c}   &He II\tablenotemark{b}&
He II/N II\tablenotemark{c}\\
\\
274-304 & 528.3\tablenotemark{d}& $0.241\pm 4.8$\%& 541.4& $0.254\pm 4.8$\%\\
243-273 & 597.6& $0.265\pm 4.6$\%& 644.8& $0.289\pm 4.5$\%\\
212-242 & 689.8& $0.276\pm 4.3$\%& 689.2& $0.279\pm 4.3$\%\\
181-211 & 777.7& $0.285\pm 4.1$\%& 876.0& $0.334\pm 3.9$\%\\
150-180 & 920.9& $0.323\pm 3.8$\%& 1000.3& $0.343\pm 3.7$\%\\
119-149 & 1151.6& $0.357\pm 3.4$\%& 1099.2& $0.343\pm 3.0$\%\\
88-118 & 1490.1& $0.389\pm 3.1$\%& 1564.9& $0.391\pm 3.0$\%\\
57-87 & 1818.4& $0.464\pm 2.8$\%& 1755.8& $0.477\pm 2.9$\%\\
\end{tabular}

\tablenotetext{a}{Position above limb in arcseconds.}
\tablenotetext{b}{Measured total counts in the He II 1085 \AA\  line.}
\tablenotetext{c}{Ratio of total counts in He II 1085 \AA\  line to total
counts in the three N II features.}
\tablenotetext{d}{Fluxes in photons s$^{-1}$cm$^{-2}$sr$^{-1}$ are given by counts
$\times 3.183\times 10^6$.}
\end{table}

\begin{table}
\caption{He II 992 \AA\ Intensities. \label{tab2}}
\begin{tabular}{rrrrr}
\\
position\tablenotemark{a} & He II\tablenotemark{b}&
He II/N III\tablenotemark{c}   &He II\tablenotemark{b}&
He II/N III\tablenotemark{c}\\
\\
274-304 & 173.7\tablenotemark{d}& $0.104\pm 8.0$\%& 140.6& $0.095\pm 8.8$\%\\
243-273 & 133.1& $0.077\pm 9.0$\%& 137.5& $0.088\pm 8.9$\%\\
212-242 & 173.7& $0.096\pm 7.9$\%& 146.1& $0.084\pm 8.6$\%\\
181-211 & 222.9& $0.107\pm 7.1$\%& 185.8& $0.104\pm 7.7$\%\\
150-180 & 236.9& $0.113\pm 6.9$\%& 259.9& $0.120\pm 6.6$\%\\
119-149 & 310.1& $0.125\pm 6.0$\%& 300.3& $0.134\pm 6.1$\%\\
88-118 & 468.0& $0.157\pm 5.0$\%& 429.6& $0.154\pm 5.2$\%\\
57-87 & 357.5& $0.142\pm 5.7$\%& 394.3& $0.179\pm 5.4$\%\\
\end{tabular}

\tablenotetext{a}{Position above limb in arcsconds.}
\tablenotetext{b}{Measured total counts in the He II 992 \AA\  line.}
\tablenotetext{c}{Ratio of total counts in He II 992 \AA\  line to total
counts in the N III 991.57 \AA\  line.}
\tablenotetext{d}{Fluxes in photons s$^{-1}$cm$^{-2}$sr$^{-1}$ are given by counts
$\times 2.911\times 10^6$.}
\end{table}

\begin{table}
\caption{O VI 1031.93 and 1037.60 \AA\ Intensities. \label{tab3}}
\begin{tabular}{rrrrrr}
\\
position\tablenotemark{a} & O VI 1031.93 \AA\ &
O VI 1037.60 \AA\    &Ratio, $R$&
$\left({2R-4\over R}\right)$\tablenotemark{b}&
$I\left({\rm He II 1085}\right)\over I\left({\rm O VII 1031.93}\right)$\\
\\
274-304 & 58588\tablenotemark{c}& 26273\tablenotemark{c}& $2.230\pm 0.74$\%& 0.206&\\
243-273 & 87624& 37619& $2.329\pm 0.62$\%& 0.283&\\
212-242 & 128668& 56892& $2.262\pm 0.50$\%& 0.232&\\
181-211 & 212679& 94810& $2.243\pm 0.39$\%& 0.217&\\
150-180 & 366531& 161865& $2.264\pm 0.30$\%& $0.233\pm 2$\%
&$8.70\times 10^{-4}\pm 13$\%\\
119-149 & 577583& 254678& $2.268\pm 0.24$\%& $0.236\pm 1.7$\%
&$7.53\times 10^{-4}\pm 10$\%\\
88-118 & 922039& 436193& $2.114\pm 0.18$\%& $0.108\pm 3.3$\%
&$7.03\times 10^{-4}\pm 7$\%\\
57-87 & 977680& 451582& $2.165\pm 0.18$\%& 0.152&\\
\end{tabular}
\tablenotetext{a}{Position above limb in arcseconds.}
\tablenotetext{b}{Fraction of 1031.93 \AA\  intensity resulting from
radiative excitation.}
\tablenotetext{c}{Fluxes in photons s$^{-1}$cm$^{-2}$sr$^{-1}$ are given by counts
$\times 2.889\times 10^6$ for 1031.93 \AA\  and ....}
\end{table}

\begin{table}
\caption{Theoretical He II Emissivities\tablenotemark{a}}
\begin{tabular}{rrrr}
\\
$\log\left(T\right)$& He II 1640\AA\ & He II 1085\AA\ & He II 992\AA\ \\
\\
5.4 & 7.46(-15)\tablenotemark{b}& 5.66(-16)& 1.79(-16)\\
5.5 & 5.39(-15)& 4.22(-16)& 1.33(-16)\\
5.6 & 3.34(-15)& 2.76(-16)& 8.65(-17)\\
5.7 & 2.34(-15)& 2.01(-16)& 6.31(-17)\\
5.8 & 1.66(-15)& 1.48(-16)& 4.64(-17)\\
5.9 & 1.18(-15)& 1.08(-16)& 3.40(-17)\\
6.0 & 8.68(-16)& 8.12(-17)& 2.55(-17)\\
6.1 & 6.34(-16)& 6.04(-17)& 1.90(-17)\\
6.2 & 4.60(-16)& 4.45(-17)& 1.40(-17)\\
6.3 & 3.40(-16)& 3.33(-17)& 1.04(-17)\\
6.4 & 2.49(-16)& 2.48(-17)& 7.76(-18)\\
\end{tabular}
\tablenotetext{a}{Emissivities are in photons/H atom/s/unit
electron density, evaluated for photospheric abundances and an
electron density of $10^8$ cm$^{-3}$.}
\tablenotetext{b}{The notation $a\left(-b\right)$ means $a^{-b}$.}
\end{table}
\begin{table}
\caption{Theoretical H I Emissivities\tablenotemark{a}}
\begin{tabular}{rrrr}
\\
$\log\left(T\right)$& H I 1215\AA\ & H I 1025\AA\ & H I 972\AA\ \\
\\
5.4 & 5.95(-14)\tablenotemark{b}& 7.06(-15)& 2.38(-15)\\
5.5 & 4.70(-14)& 5.68(-15)& 1.91(-15)\\
5.6 & 3.27(-14)& 4.05(-15)& 1.36(-15)\\
5.7 & 2.48(-14)& 3.11(-15)& 1.05(-15)\\
5.8 & 1.88(-14)& 2.37(-15)& 7.99(-16)\\
5.9 & 1.40(-14)& 1.79(-15)& 6.01(-16)\\
6.0 & 1.06(-14)& 1.36(-14)& 4.59(-16)\\
6.1 & 7.88(-15)& 1.02(-15)& 3.43(-16)\\
6.2 & 5.75(-15)& 7.47(-16)& 2.51(-16)\\
6.3 & 4.19(-15)& 5.44(-16)& 1.83(-16)\\
6.4 & 2.99(-15)& 3.88(-16)& 1.31(-16)\\
\end{tabular}
\tablenotetext{a}{Emissivities are in photons/H atom/s/unit
electron density, evaluated for photospheric abundances and an
electron density of $10^8$ cm$^{-3}$.}
\tablenotetext{b}{The notation $a\left(-b\right)$ means $a^{-b}$.}
\end{table}
\begin{table}
\caption{H I Lyman $\beta$ and Lyman $\gamma$ Intensities. \label{tab4}}
\begin{tabular}{rrrrr}
\\
remark& position\tablenotemark{a} & Lyman $\beta$ & Lyman $\gamma$& Ratio
$\beta$/$\gamma$ \\
\\
total counts & 88-118 & $56573\pm 0.4$\%\tablenotemark{c}&
$13809\pm 0.9$\%\tablenotemark{c}\\
total counts (assumed &
243-273 & $17074\pm 0.8$\%& $4209\pm 1.5$\%& $0.251\pm 1.7$\%\tablenotemark{b}\\
pure scattered light)& \\
\\
scattered light& 88-118& $31160\pm 1.8$\%& $7681\pm 4$\%\\
subtract scattered light& 88-118& $25413\pm 2.4$\%& $6127\pm 3.1$\%\\
subtract He II& 88-118& $25110\pm 2.4$\%& $6116\pm 3.1$\%&
$0.248\pm 4$\%\tablenotemark{c}\\
\end{tabular}
\tablenotetext{a}{Position above limb in arcseconds.}
\tablenotetext{b}{Ratio of disk intensities, includes correction for
instrument sensitivity.}
\tablenotetext{c}{Fluxes in photons s$^{-1}$cm$^{-2}$sr$^{-1}$ are given by counts
$\times 2.948\times 10^6$ for 972 \AA\  and $\times 2.890\times 10^6$ for 1025 \AA\ .}
\tablenotetext{d}{Ratio of coronal intensities, includes correction for
instrument sensitivity.}
\end{table}

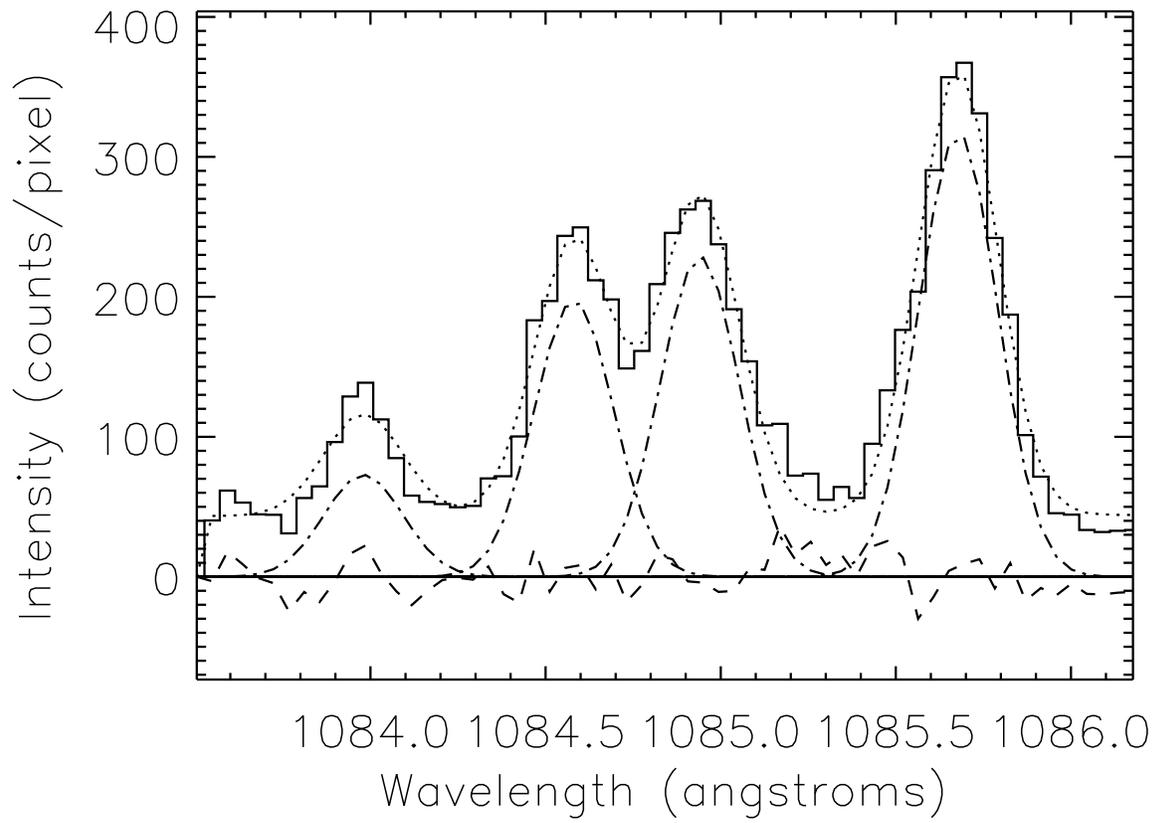
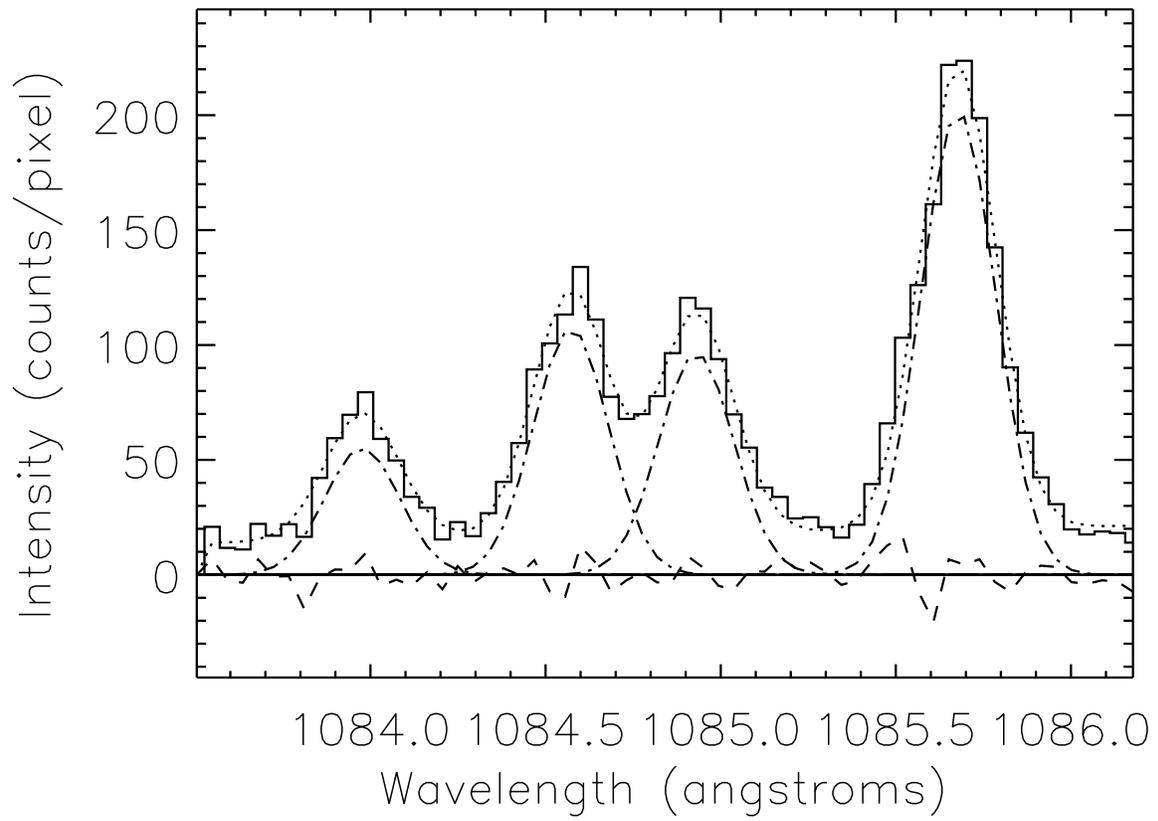
\end{document}